\documentclass{iaus}
\usepackage{graphicx}
\usepackage{natbib}

\newcommand {\siiii} {{\rm Si}\,{\small\rm III}}
\newcommand {\siiv} {{\rm Si}\,{\small\rm IV}}
\newcommand {\cii} {{\rm C}\,{\small\rm II}}

\newcommand {\civ} {{\rm C}\,{\small\rm IV}}

\newcommand {\kpc} {\,{\rm kpc}}

\newcommand {\cmmq}{\,{\rm cm^{-2}}}

\newcommand {\mo}{\,{\rm M}_\odot}

\newcommand{\Myr}{\,{\rm Myr}}
\newcommand{\Gyr}{\,{\rm Gyr}}
\newcommand{\K}{\,{\rm K}}

\newcommand {\moyr}{\,{{\rm M}_\odot\,\rm yr}^{-1}}

\newcommand {\apj}{ApJ}
\newcommand {\aj}{AJ}
\newcommand {\apjs}{ApJS}
\newcommand {\apjl}{ApJL}
\newcommand {\mnras}{MNRAS}
\newcommand {\aap}{A\&A}
\newcommand {\araa}{ARA\&A}
\newcommand {\aapr}{A\&ARv}

\newcommand {\nat}{Nature}

\title[How can star formation be sustained?] 
{How can star formation be sustained?}

\author[Filippo Fraternali]   
{Filippo Fraternali
}

\affiliation{Department of Physics and Astronomy, University of Bologna (I)\\ 
Kapteyn Astronomical Institute, University of Groningen (NL)\\ email: {\tt filippo.fraternali@unibo.it} \\[\affilskip]
}

\pubyear{2014}
\volume{298}  
\pagerange{XX -- YY}
\jname{IAUS\,298 Setting the scence for Gaia and LAMOST}
\editors{S. Feltzing, G. Zhao, N.\,A. Walton \& P.\,A. Whitelock, eds.}
\begin{document}

\maketitle

\begin{abstract}
There is overwhelming evidence that the Milky Way has formed its stars at a relatively constant rate throughout the Hubble time. 
This implies that its stock of cold gas was not in place since the beginning but it has been acquired slowly through gas accretion.
The gas accretion must have been at low metallicity in order to reconcile the metallicities observed in the disc with the predictions of chemical evolution models.
But how does this gas accretion take place?
I review the current evidence of gas accretion into the Milky Way and similar galaxies through the infall of cold gas clouds and satellites.
The conclusion from these studies is that the infalling gas at high column densities observed in HI emission is a least one order of magnitude below the value required to sustain star formation, thus accretion must come from a different channel.
The likely reservoir for gas accretion is the cosmological corona of virial-temperature gas in which every galaxy must be embedded.
At the interface between the disc and the corona the cold high-metallicity disc gas and the hot coronal medium must mix efficiently and this mixing causes the cooling and accretion of the lower corona.
I show how this mechanism reproduces the kinematics of the neutral extraplanar gas in the Milky Way and other nearby galaxies and the ionised high-velocity clouds observed in HST spectra.
I conclude with the speculation that the loss in efficiency of the disc-corona interaction is the ultimate cause for the evolution of disc galaxies towards the red sequence.

\end{abstract}

\firstsection 

\section{Introduction}

Over the last decade we have witnessed great improvements in our understanding of the large-scale structure and evolution of the Universe. 
The cosmological parameters have been measured with unprecedented precision \citep[e.g.,][]{Spergel+07}, star formation in galaxies has been traced in detail \citep[e.g.,][]{Hopkins&Beacom06} and cosmological simulations have successfully reproduced the mass assembly \citep[e.g.,][]{Springel+06}.
However, a major obstacle still divides us from a comprehensive understanding of how the visible structures in the Universe formed and evolved, i.e., understanding the role of gas.
Cosmology tells us that ordinary matter, commonly refereed to as baryons, amounts to only $\sim 5\%$ of the universal matter/energy budget \citep{Planck}.
Remarkably, the amount of baryons in collapsed structures in the Universe (galaxies and galaxy clusters) appears to be much less, of the order $0.5-0.8\%$ of the total \citep[e.g.,][]{McGaugh+10}.
In other words, most of the baryons are \emph{missing} from cosmological structures and must therefore reside in regions outside galaxies \citep{Bregman07}.

The current wisdom about missing baryons is that they are in a gaseous form and permeate the intergalactic medium (IGM).
A large fraction (around 30\%) of these baryons can be at relatively low temperatures (around few $\times 10^4 \K$) and form the local Lyman-$\alpha$ forest \citep{Penton+04}.
The rest is likely to be in a rarefied medium at higher temperatures (typically from $10^5$ to $10^7 \K$) located in intergalactic space \citep{Shull+12}.
The quest for this warm-hot intergalactic medium (WHIM) is still ongoing as its detection is at the limit of current X-ray facilities \citep[e.g.,][]{Nicastro+05, Kaastra+06}.
In the regions close to galaxies the WHIM should be much easier to observe as it forms hot atmospheres that reach column densities and emissivities detectable with the available satellites \citep[e.g.,][]{Crain+10, Anderson&Bregman10}.
These atmospheres ({\it cosmological} coronae) are expected to contain a large fraction of the baryons associated to a galactic potential well \citep{Fukugita&Peebles06} and yet their detection proved elusive \citep{Rasmussen+09}.
In the last couple of years the first convincing detections were obtained, all coming from massive late-type galaxies: NGC\,1961 \citep{Anderson&Bregman11, Bogdan+12}, NGC\,12591 \citep{Dai+12}, and NGC\,6753 \citep{Bogdan+12}.
The typical extent of these coronae are $\sim50-80 \kpc$ from the centre of the galaxy and the total masses are calculated by extrapolating to the virial radius, with obvious uncertainties.
The general result is that the amount of baryons contained in these coronae is limited and close to $\sim 10\%$ of the value required to explain all missing baryons.

The above results agree nicely with recent indirect measures of the corona of the Milky Way \citep{Anderson&Bregman10}.
\citet{Gatto+13}, using ram-pressure stripping from dwarf galaxies around the Milky Way estimated that the mass of the Milky Way's corona could be around $10-20\%$ of the baryons missing from the Galaxy potential well.
It is worth noticing that even $10-20\%$ of the expected value makes up a mass of the corona at least as large as the mass of the entire stellar disc, thus the corona is potentially a huge reservoir of gas if it can be used for star formation.
Here, I will describe a mechanism that can ensure this vital process for the life of the Galaxy.
I first justify the need for gas accretion during the building up of the Milky Way's stellar disc (Section \ref{sec:case}).
Then I describe the current evidence for gas accretion (Sections \ref{sec:inventory} and \ref{sec:corona}).
Finally, I present the likely mechanism that allows the disc to extract gas from the corona (Section \ref{sec:refrigerator}) and its observational evidence (Section \ref{sec:evidence}).
I conclude with some considerations about the evolution of the Milky Way and other late-type galaxies (Section \ref{sec:conclusions}).

\section{The case for gas accretion}
\label{sec:case}

\begin{figure}[ht]
\begin{center}
 \includegraphics[width=\textwidth]{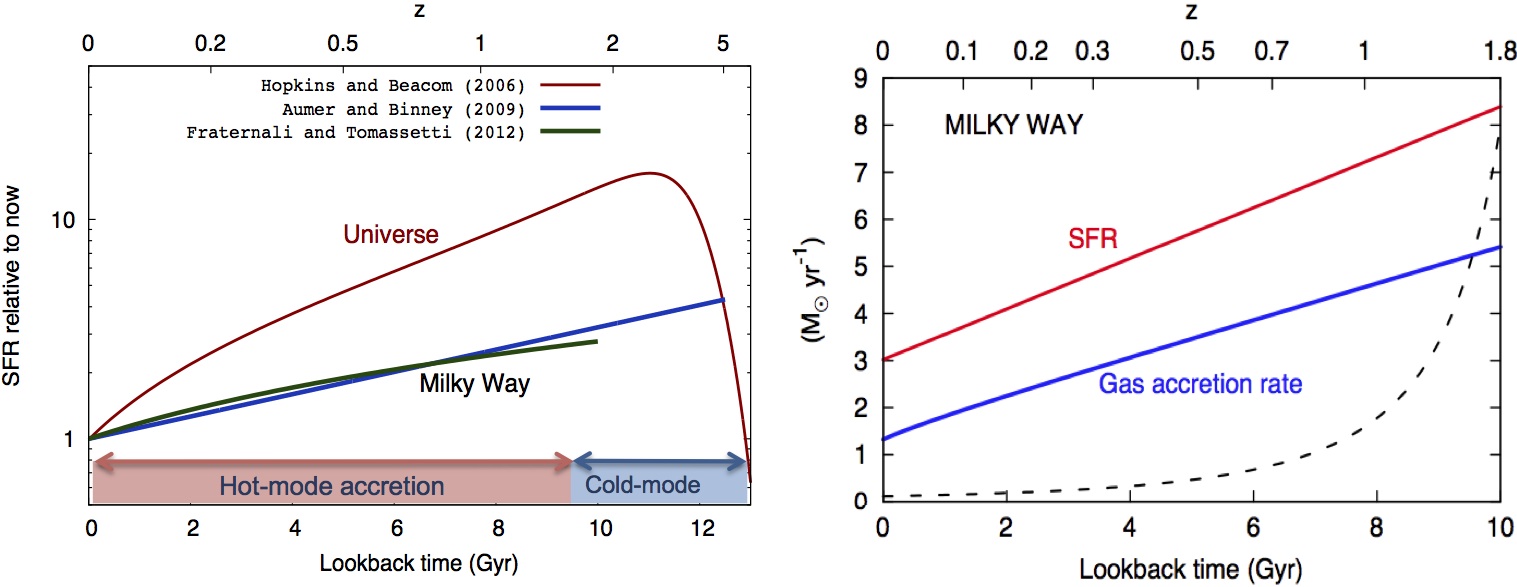} 
 \caption{{\it Left:} comparison between two determinations of the SFH of the Milky Way \citep{Aumer&Binney09, Fraternali&Tomassetti12} and the average star formation rate density of the Universe \citep{Hopkins&Beacom06}.
The three distributions are normalized at the current time.
{\it Right}: reconstruction of the SFH of the Milky Way's disc and the gas accretion rate required by the Kennicutt-Schmidt law \citep[see][]{Fraternali&Tomassetti12}; the dashed line shows the evolution of a closed-box galaxy starting with the same initial amount of gas.
}
\label{FF:fig:sfhs} 
\end{center}
\end{figure}

There is a striking difference between the evolution of the star formation rate (SFR) density in the Universe and the typical star formation history (SFH) of a galaxy like the Milky Way (Figure \ref{FF:fig:sfhs}, left panel). 
The SFH of our Galaxy has been derived by several authors \citep[e.g.,][]{Twarog80,Rocha-Pinto+00}, and in Fig.\ \ref{FF:fig:sfhs} I show two recent determinations.
If one normalizes the Universal and Galactic SFR at the current time it is clear that the shapes of the two functions differ dramatically by up to a factor of $\geq 5$ at $z\sim2$.
This shows that different types of galaxies contribute in a very different way to the Universal star formation.
A breakup of the star formation density in bins of galaxies with different stellar masses shows that most early type massive galaxies formed stars at early epochs \citep[e.g.,][]{Panter+07} and the rise of the SFR density of the Universe around $z=2$ is dominated by galaxies that moved quickly to the red sequence \citep[e.g.,][]{Pipino+13}.
On the other hand, blue galaxies such as the Milky Way, residing in the so-called main sequence \citep[e.g.,][]{Elbaz+11} have formed stars throughout the Hubble time at a relatively constant rate.

The right panel of Fig.\ \ref{FF:fig:sfhs} shows a reconstruction of SFH and gas accretion in the Milky Way, the SFH is the same of the left panel \citep{Fraternali&Tomassetti12}.
The gas accretion is estimated by imposing the Kennicutt-Schmidt law \citep{Schmidt59, Kennicutt98a}.
This simple requirement generates a reciprocal and almost constant ratio between gas accretion and star formation rate at every time as the gas mass in the disc remains nearly constant.
All this also implies that the large majority of stars in the Milky Way's disc have formed after $z=1$ \citep[e.g.,][]{Aumer&Binney09} and leads to a potential paradox.
As shown in the left panel of Fig.\ \ref{FF:fig:sfhs}, according to recent cosmological simulations galaxies go through two phases of gas accretion \citep[e.g.,][]{Dekel+09}.
In the first, cold-mode phase, the gas falls into the potential well without being shock heated to the virial temperature and therefore reaches very quickly the bottom of the potential well \citep{Binney77}.
Eventually, this cold-mode accretion is substituted by a hot-mode phase that produces and feeds a rarefied hot atmosphere around the Galaxy (see Section \ref{sec:corona}).
The transition between these two phases occurs around $z=1-2$ for a Milky-Way--like galaxy \citep{Keres+09}.
The paradox is that most of the star formation and therefore most of the {\it cold} accretion must take place during the hot-mode phase.
Recent moving-mesh grid simulations with AREPO make this problem even more severe \citep{Nelson+13}.
I discuss a possible solution to this paradox in Section \ref{sec:refrigerator}.

The need for continuous gas accretion in the disc of the Milky Way can also be made from completely independent arguments.
The determination of stellar ages show that the thin and thick disc stars have formed rather uniformly between now and $\sim 12$ Gyrs ago \citep{Bensby&Feltzing10}.
Chemical evolution models cannot reproduce the abundances observed in stars or the abundance gradients without extensive and continuous accretion of fresh unpolluted gas \citep[e.g.,][]{Chiappini+01, Schoenrich&Binney09}.
Finally, the amount of deuterium observed in the {\it local} interstellar medium (ISM) becomes consistent with the primordial abundance only considering conspicuous accretion of pristine material onto the disc \citep[e.g.,][]{Steigman+07}.
In conclusion, all these investigations show that the Milky Way (as presumably any blue-cloud galaxy) must have continuously accreted {\it nearly} primordial gas and mixed it with the reprocessed ISM to form subsequent generations of stars.
In the next Section, I make a critical investigation of the amount of gas accretion that is currently observed.

\section{Inventory of gas accretion in the Local Universe}
\label{sec:inventory}

One way for disc galaxies to acquire gas is through interactions with gas-rich satellites that can lose their cold ISM via tidal and ram pressure stripping \citep[e.g.,][]{Mayer+06}.
\citet{Sancisi+08} estimated the accretion from satellites by looking for disturbances in the outer HI discs of local galaxies and found an accretion rate of the order $0.1-0.2 \moyr$.
We have recently built an automatic algorithm to detect HI-rich companions around galaxies and estimated a firm upper limit to the accretion rate (Di Teodoro \& Fraternali, in prep.).
We made three very conservative assumptions: i) all companions will merge, ii) they will merge in the shortest possible time, and iii) once the companion reaches the main galaxy all its gas is transferred instantaneously.
Despite these extreme assumptions we found that the accretion rate contributed by satellites can at most be $0.23 \moyr$, which is six times lower that the average SFR of our sample.
The actual value is likely to be much (orders of magnitude) lower than this, thus minor merges can not feed star formation in local galaxies.

The presence of High-Velocity Clouds (HVCs) falling towards the disc of the Milky Way \citep{Wakker&vanWoerden97} was established soon after the first surveys of the Northern sky in the 21-cm line \citep{Muller+63}.
For decades the HVCs were regarded as the gas supply for the Milky Way's star formation even though their origin was much debated \citep[e.g.,][]{Oort70, Bregman80, Blitz+99}.
In the last ten years or so, precise distances to the main clouds have been determined using halo stars to bracket their positions with respect to us \citep{Wakker+08, Thom+08}.
The results showed that most of the HVCs are at about $10 \kpc$ from us and thus they are not very massive (with typical $M_{\rm HI}\leq10^6 \mo$).
A recent determination of the accretion rate that can come from HVCs led to a mere $\dot M_{\rm acc}=0.08 \moyr$, including Helium and assuming that there is as much ionised gas as is visible in HI \citep{Putman+12}.
Clearly, HVCs fall more than order of magnitude too short to feed the star formation in the Milky Way.
Here I assume a current Galactic SFR of $\simeq2.0 \moyr$ \citep{Chomiuk&Povich11}.

\begin{figure}[ht]
\begin{center}
 \includegraphics[width=\textwidth]{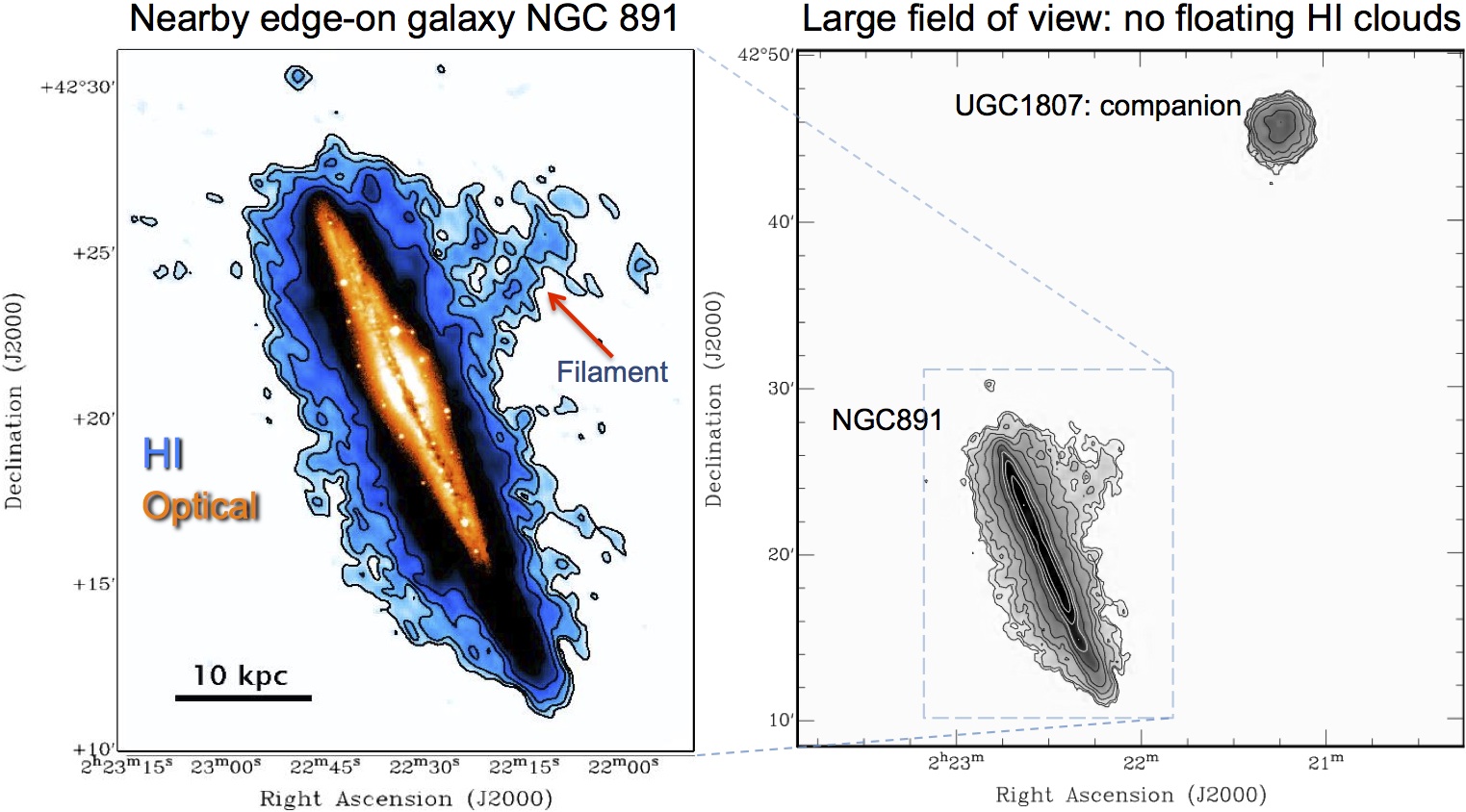}
 \caption{Total HI map of NGC\,891 from the ultra-deep observations of \citet{Oosterloo+07}.
\emph{Left}: the optical disc is surrounded by an HI halo extending everywhere up to $8-10 \kpc$ from the plane, with a long filament (labelled) reaching almost $20 \kpc$.
\emph{Right}: the wealth of HI features observed in proximity to the disc of NGC\,891 contrasts with the emptiness of the large field where no clouds are detected down to a mass limit of a few times $10^5 \mo$.}
   \label{FF:fig:n891} 
\end{center}
\end{figure}

The presence of neutral gas around galaxies has also been studied in detail in external galaxies.
The problem with these observations is that they require very long integration times to get to the column densities (typically below $10^{20} \cmmq$) needed to detect the extraplanar component \citep{Swaters+97, Fraternali+01}.
Several observations now reach the required column densities, notably the observations of M31 \citep[e.g.,][]{Thilker+04} and NGC\,891 \citep{Oosterloo+07}.
The first may have some issues related to the proximity in velocity to the Milky Way's halo gas \citep{Marasco&Fraternali11}, the Magellanic Stream \citep{Putman+03} and the interaction with M33 \citep[e.g.,][]{Bekki08}.
NGC\,891 is instead a clean case.
This perfectly edge-on galaxy is located 9.5 Mpc away from us and has a systemic velocity incompatible with Milky Way's HVCs.
It has been observed in a campaign carried out with the Westerbork Synthesis Radio Telescope for more than 200 hours, and is one of the deepest HI observations ever made.
The total HI map resulting from this effort is shown in Fig.\ \ref{FF:fig:n891}, left panel.
A large extraplanar component has been detected amounting to about $25\%$ of the total HI mass, or $1.2\times 10^{9} \mo$.
Most of this gas is located within $\sim 8 \kpc$ from the disc with a filament extending up to about $20 \kpc$.
Although we now regard NGC\,891 as a remarkable and likely extreme case of the phenomenon of extraplanar gas, several other galaxies observed with the right sensitivity also show the presence of this component \cite[e.g.,][]{Fraternali+02, Barbieri+05, Heald+11}.
In particular galaxies with SFRs of order $\geq 1 \moyr$ seem to contain substantial amounts of extraplanar gas \citep[e.g.,][]{Boomsma+08, Marasco&Fraternali11, Gentile+13}.
This fact together with the success of galactic fountain models in reproducing the extraplanar gas \citep{Fraternali&Binney06,Fraternali&Binney08} shows that the vast majority of extraplanar material must come from the disc through supernova feedback.
This has also been confirmed by the determination of the metallicity of the gas above the plane in NGC\,891 that is about Solar \citep{Bregman+13}.

But where is the gas accretion?
In section \ref{sec:refrigerator}, I will argue that the gas accretion is taking place in the extraplanar medium but it is hidden from direct detection being mixed with the galactic-fountain cycle.
Here, I stress the point already made about the HVCs of the Milky Way.
If we take a larger view of the field around NGC\,891 (right panel of Fig.\ \ref{FF:fig:n891}) we discover two important facts.
There is a companion (UGC\,1807) at a projected distance from NGC\,891 of $80 \kpc$, and located at the linear extrapolation of the filament.
\citet{Mapelli+08} actually explain this filament as the product of an interaction with the companion.
The second, more remarkable fact is that around NGC\,891, beyond $10-20 \kpc$ there is basically no emission detected anywhere.
The sensitivity of the instrument drops significantly only beyond $\sim 20'\simeq 60 \kpc$.
Thus if HI clouds with masses of a few $\times 10^5 \mo$ or larger existed between $10 \kpc$ and $60 \kpc$ from the disc we would have detected them.
This lack of floating clouds is confirmed by several other studies of nearby galaxy groups \citep[e.g.][]{Pisano+07, Chynoweth+09} and suggests that most Galactic HVCs, including the compact ones, are likely nearby the Milky Way's disc; the notable exception being the Magellanic Stream and the plethora of shredded clouds that surrounds it \citep{Putman+03}.

In addition to HI in emission, accretion can also be observed in absorption using Ly$\alpha$ or metal lines \citep[e.g.,][]{Rubin+12}.
In the last years, a series of features produced by ionised species (e.g., \siiii, \siiv, \cii, \civ) have been observed in Hubble Space Telescope (HST) spectra \citep[e.g.,][]{Shull+09, Collins+09, Lehner+12}.
The accretion rates derived from these features are quite uncertain but they seem to get closer to the required $\sim 1 \moyr$.
They clearly show material at intermediate temperatures between the Galactic corona and the cold disc. 
I will return to this point in Section \ref{sec:evidence} where I will show that several of these ionised HVCs are likely to be very close to the disc \citep{Lehner&Howk11}.
Interestingly, material at similar temperatures is also observed further out in Ly$\alpha$ absorption but in exactly the same amount in early and late type galaxies, suggesting that this circumgalactic medium does not take part in gas accretion \citep[e.g.,][]{Thom+12}.

\section{The cosmological corona}
\label{sec:corona} 

As already mentioned, it is a strong prediction of models of structure formation in the Universe that galaxies should be embedded in virial-temperature atmospheres, called cosmological coronae \citep[e.g.,][]{White&Rees78, Fukugita&Peebles06}.
Until a few years ago it was thought that coronae of disc galaxies could easily cool to produce cold clouds that would fall into the disc and feed its star formation \citep[e.g.,][]{Kaufmann+06}.
This idea was challenged recently by analytical calculations and hydrodynamical simulations showing that thermal instabilities cannot develop in the stratified corona typical of a galaxy like the Milky Way \citep{Binney+09, Joung+12}.
This is qualitatively in agreement with the fact that early type galaxies, despite having prominent X-ray coronae \citep{Forman+85} do not have significant amounts of cold gas \citep{Morganti+06} and consequently star formation.
This may be even more surprising if one considers that the metallicity of the coronal gas of early type galaxies is close to Solar \citep{Ji+09} showing that a large fraction of this gas is contributed by feedback from evolved stellar populations \citep[e.g.,][]{Ciotti&Ostriker97}.
In early type galaxies the cooling of the corona may occur in the very centre where the densities are higher but this will feed the central black-hole, causing feedback and re-heating of the corona \citep{Binney&Fraternali12}.

The situation for disc galaxies is instead dramatically different.
The metallicity of the corona is likely low and the cooling time is very long, of order of tens of Gyrs reaching perhaps a few Gyrs close to the disc \citep{Hodges-Kluck&Bregman13}.
Nevertheless, these galaxies are somehow capable of extracting cold gas from their coronae at a high rate and in a large area, covering the whole stellar disc.
All the mechanisms that are usually invoked for this (e.g., thermal instabilities or cold filaments penetrating the corona, \citep{Fernandez+12}) should also be effective in early type galaxies and yet they are not.
In other words, the cooling seems to occur only in coronae where a disc of cold star-forming gas is already present.
In the following I describe the mechanism that can solve this apparent contradiction.

\section{The refrigerator of the corona}
\label{sec:refrigerator}

From the previous sections we can draw the following conclusions:
(i) star forming galaxies require the continuous accretion of fresh gas to feed their star formation;
(ii) the available gas supply is the virial-temperature corona;
(iii) the corona cools efficiently only in disc galaxies and not in early type galaxies.

In a series of papers starting from \citet{Fraternali&Binney08} we have been suggesting that the likely mechanism to extract star forming gas from the corona is its pollution with gas coming from the star-forming disc.
This pollution happens because supernova feedback in the disc is continuously ejecting high metallicity gas into the halo region through a galactic fountain \citep[e.g.,][]{Houck&Bregman90} and occasionally winds \citep[e.g.,][]{Oppenheimer+10}.
This gas is mostly cold \citep{Melioli+08} and it is bound to interact with the low-metallicity coronal gas.  
In the region of interface the mixing of the two media reduces the cooling time of the corona by orders of magnitudes causing it to condensate into fountain clouds and fall back to the disc \citep{Marinacci+10}.
Most of this interaction is likely to happen in a region close to the disc (a few kpc) where the galactic fountain operates and the column density of the fountain gas is the highest (see Section \ref{sec:inventory}).

\begin{figure}[ht]
\begin{center}
 \includegraphics[width=\textwidth]{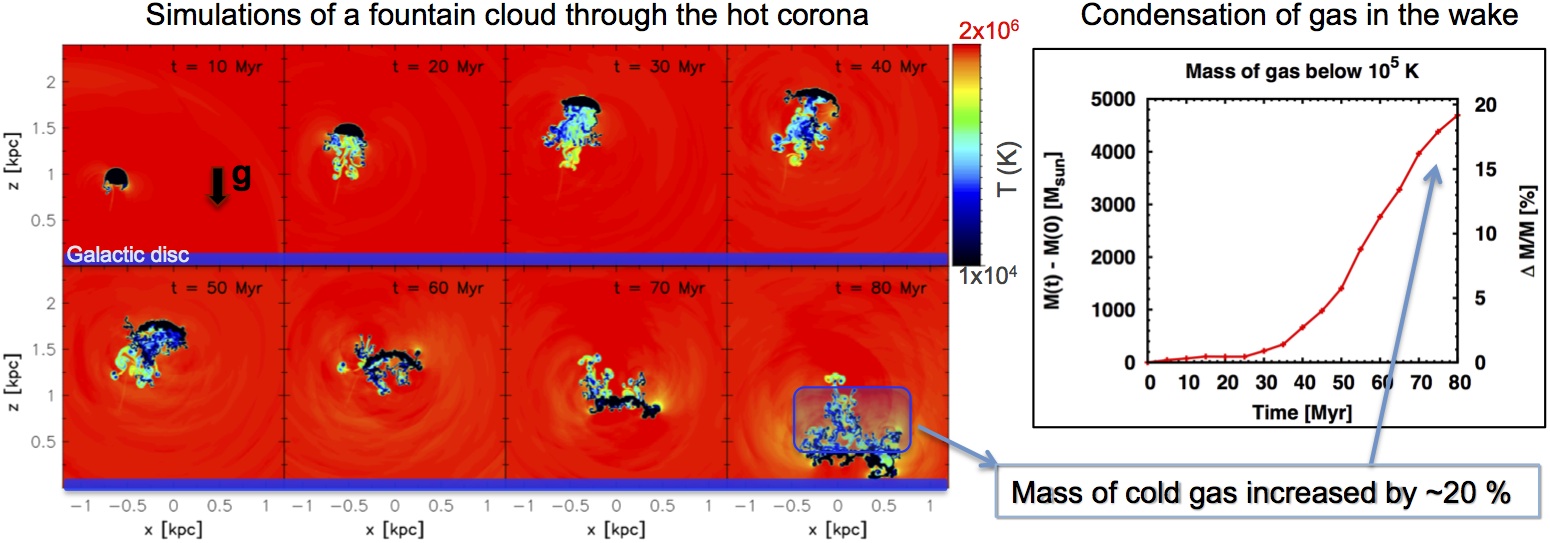}
 \caption{Hydrodynamical simulation (temperature snapshots at different times, indicated in the top right corner of each frame) of a high-metallicity fountain cloud leaving the star-forming disc and interacting with the low-metallicity corona. 
In the wake of the cloud part of the corona condenses into cold clumps that fall back to the disc.
At the end of the simulation the amount of cold gas increases by about $20\%$ for parameters typical of the Milky Way clouds and corona.}
   \label{FF:fig:sim} 
\end{center}
\end{figure}

The mechanism of supernova-driven cooling of the corona has been studied with high (typically 1 pc) resolution hydrodynamical simulations like the one shown in Fig.\ \ref{FF:fig:sim}.
The cold cloud ejected from the disc suffers Kelvin-Helmholtz instability and an extended wake forms behind it.
In the wake, regions of condensation develop and the amount of cold gas, despite ablation, increases with time (right panel).
The material condensing from the corona is then accreted onto the disc and can feed new star formation.
The exact amount of condensation that is produced in these simulations depends critically on two parameters: the temperature and density of the corona. 
For instance, the higher the temperature the lower the condensation. 
Although this has not been fully explored yet, there is presumably a temperature threshold beyond which condensation does not occur and this may cause the mechanism to stop working at large halo masses (large virial temperatures), see Section \ref{sec:conclusions}.
Also, the higher the coronal density the larger the condensation, this is essentially due to the increasing stripping and mixing in the presence of high densities.
Finally, the metallicity of the cloud also seems important, and experiments with clouds at half Solar metallicity show significant reduction in the condensation (Armillotta, Fraternali \& Marinacci, in prep.).

\section{Observational evidence of supernova-driven cooling}
\label{sec:evidence}

The model described above can easily be tested using observations of extraplanar gas in nearby disc galaxies.
In a non-interacting galactic fountain the trajectories of the clouds expelled from the disc are given only by the gravitational potential that is well constrained by the HI rotation curve \citep{Fraternali&Binney06}.
When the fountain clouds interact with the corona, the trajectories are significantly modified, and it is the signature of this modification that we can observe to test the validity of the model.
Moreover, the larger the condensation of the corona the more prominent the modification of the trajectories will be thus we can use the kinematics of the extraplanar gas not only to detect gas accretion but also to quantify its amount.
The first success of this idea was reported in the original paper \citep{Fraternali&Binney08} where we showed that a non-interacting galactic fountain could not reproduce the kinematics of the extraplanar HI in NGC\,891 and NGC\,2403 and only by making the fountain interact with an external medium at low angular momentum could we reconcile it with the data \citep[see also][]{Heald+07}.
The subsequent investigation carried out with hydrodynamical simulations allowed us to improve our galactic fountain model to include realistic kinematics for the corona and a new parametrization of the condensation \citep{Marinacci+10, Marinacci+11}.
\citet{Marasco+12} used this improved model to compare our predictions with the data of the Milky Way from the LAB HI survey \citep{Kalberla+05}.
A careful minimization of the residuals (along each line of sight) between the models and the whole datacube showed that also these data required condensation of the corona onto fountain clouds at a rate of $\dot M_{\rm acc}=2 \moyr$. 
Thus, the kinematics of extraplanar HI in the Milky Way is fully consistent with the accretion of gas from the corona (refrigerated by the interaction with the fountain) at a rate very similar to the Galactic SFR.

\begin{figure}[ht]
\begin{center}
 \includegraphics[width=\textwidth]{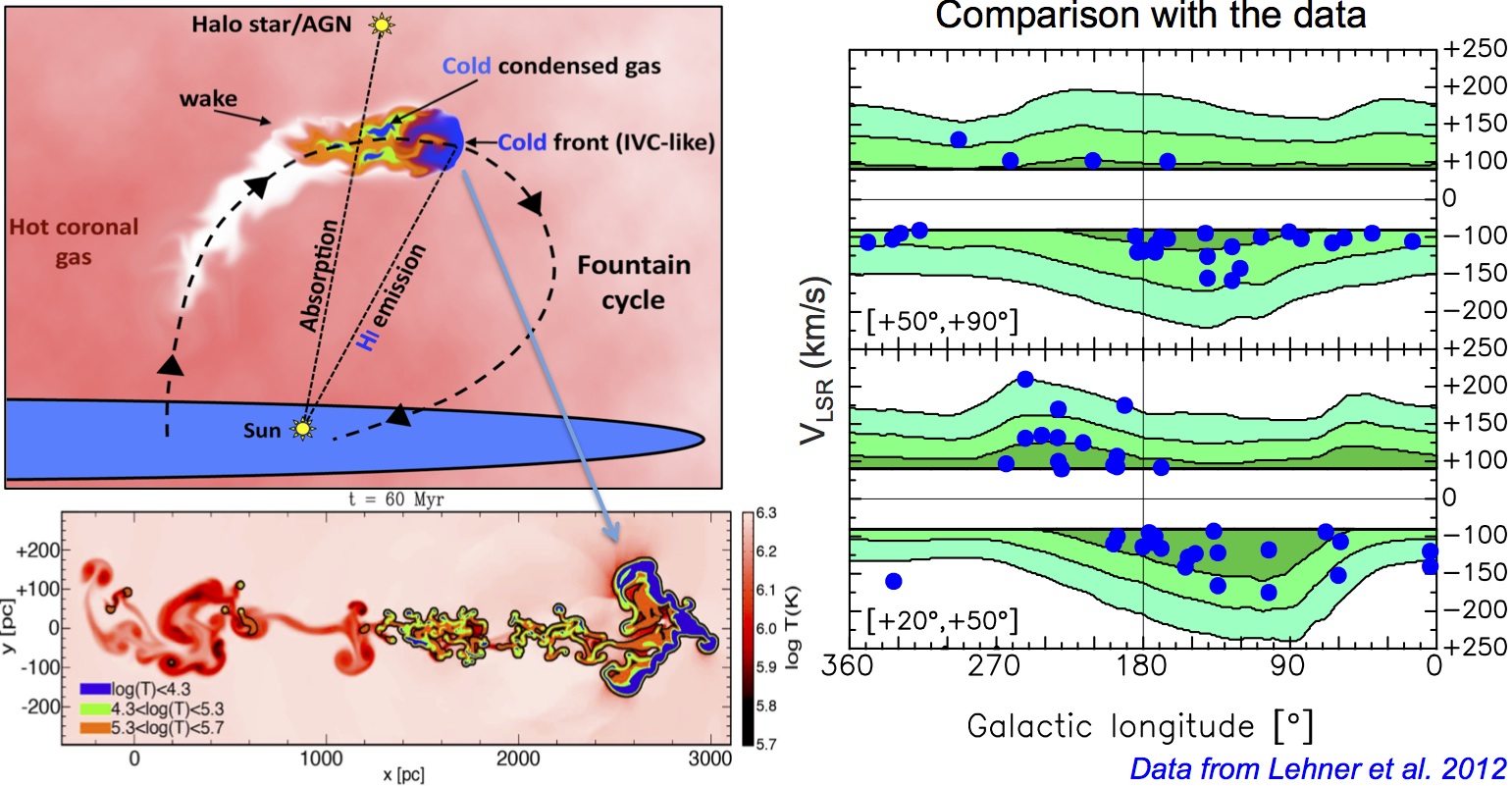}
 \caption{\emph{Top left}: schematic representation of a galactic fountain cloud interacting with the ambient hot corona. \emph{Bottom left}: hydrodynamical simulation (temperature snapshot at $t=60 \Myr$) of the cloud, outlining the different temperature ranges present in the wake. \emph{Right}: prediction for the locations and velocities of the warm ($4.3<\log(T)<5.3$) gas in our galactic fountain model compared to the detections of ionised warm (Si III, Si IV, C II, C IV) absorbers from \citet{Lehner+12}.}
 \label{FF:fig:ionised} 
\end{center}
\end{figure}

During the last year, we have extended the comparison to a completely different dataset \citep{Marasco+13}.
The idea is shown in Fig.\ \ref{FF:fig:ionised}.
The top left panel shows a cloud travelling through the halo and producing a wake, where material from the corona can cool down from the virial temperature to lower and lower temperatures encompassing the typical temperatures of the ionised species observed in the HST spectra \citep{Shull+09, Lehner+12}.
We can then take our simulations and isolate only the gas at these temperatures (bottom left panel).
Our fit of the LAB data describes the kinematics of the cold (HI) gas, the relation between cold and warm gas is simply given by the simulations, and we can therefore predict the location of the warm gas in the position-velocity space along each line of sight.
Note that in doing so we are left with no free parameters as everything is set either by the fit to the HI datacube or by the hydrodynamical simulations.
The right panel of Fig.\ \ref{FF:fig:ionised} shows the prediction of our model for where the absorbers should most likely lie (darker color means more probable).
The data (dots) are taken from \citet{Lehner+12}.
A statistical test shows that 95\% of the absorbers are consistent with our model, and so are their column densities and the number of clouds intersected by the lines of sight \citep{Marasco+13, Fraternali+13}.
In conclusion, it appears that most of the ionised HVCs are formed by the cooling of the corona triggered by the passage of disc material ejected by supernova feedback.
The amount of accretion that we estimate for this component is about $1 \moyr$ fully consistent with the estimates of \citet{Shull+09} and \citet{Lehner&Howk11}.

\section{Concluding remarks}
\label{sec:conclusions}

In the previous section we showed how the refrigerator (supernova-driven cooling) mechanism reproduces the data of local galaxies in great detail.
In particular it explains the kinematics of the extraplanar HI both in external galaxies and in the Milky Way \citep{Fraternali&Binney08, Marasco+12}.
Moreover, the same model that fits the HI data, also explains the vast majority of ionised warm absorbers observed with HST around the Milky Way \citep{Lehner+12, Marasco+13}.
According to theory, disc galaxies must have been surrounded by massive hot coronae for most of their lives \citep[e.g.,][]{Keres+09}.
It is therefore conceivable that the refrigerator mechanism has also had an important role in their evolution in the past.

\begin{figure}[ht]
\begin{center}
 \includegraphics[width=\textwidth]{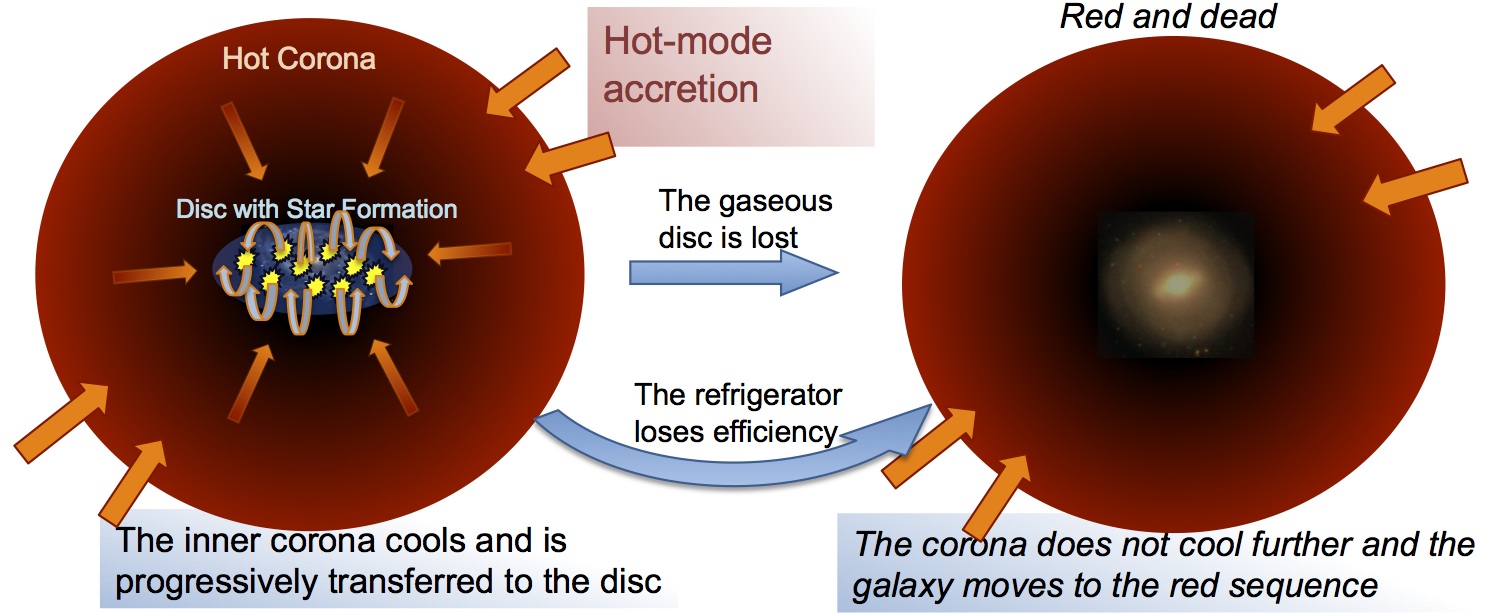}
 \caption{A schematic representation of the system cold disc $+$ cosmological corona in a star-forming galaxy and the possible evolution to the red sequence (\emph{right}) when the disc is lost as a consequence of a major merger (rapid quenching) or by loss of accretion efficiency (slow quenching).}
   \label{FF:fig:evolution} 
\end{center}
\end{figure}

In Fig.\ \ref{FF:fig:evolution}, I present a possible evolution scheme.
The galaxy with a cold gaseous disc is sitting at the bottom of a large cosmological hot corona (not to scale).
Star formation in the disc produces high metallicity gas and stellar feedback ejects part of this material from the disc bringing it into contact with the low-metallicity corona.
The mechanism described in Section \ref{sec:refrigerator} causes the cooling of the lower corona and its accretion onto the disc; the star-forming disc is effectively carving the corona from within.
In a steady state, the loss of pressure from the inner corona is compensated for by new material coming from the outer regions. This part of the process has however yet to be explored theoretically.
The outer corona is effectively fed by hot-mode accretion providing a nearly unlimited supply of gas.
As long as the conditions for condensation, briefly discussed at the end of Section \ref{sec:refrigerator}, are satisfied, the disc can keep extracting material from its corona.
In cases when the process becomes less efficient (e.g., the temperature of the lower corona is too high) the disc will exhaust the available gas in a relatively short time \citep{Lilly+13}.
If the disc is lost more rapidly (e.g., a major merger occurs) the star formation in the galaxy will be quickly quenched.
In both cases, the galaxy will move to the red sequence keeping its corona, but without a star forming disc it will have lost the ability to cool the corona efficiently, it will lose its cold gas supply and become \emph{red and dead}.

\vspace{0.5 cm}
{\bf Acknowledgements}
I acknowledge financial support from LKBF (subsidy 13.1.041), IAU, and PRIN MIUR 2010-2011, prot.\ 2010LY5N2T.


\begin{discussion}

\discuss{Ning-Chen Sun}{Why do we need some source to sustain star formation in the Milky Way as there is much gas in it? Is there any evidence that star formation will continue for longer than gas mass/SFR?}

\discuss{Answer}{The amount of cold gas in the Milky Way's disc today is actually scant and sufficient for star formation to proceed only for about $2 \Gyr$ (gas consumption timescale). Other disc galaxies in the local Universe are in the same situation. The fact that the SFR has been constant in the past shows that also the gas consumption timescale was the same, so for instance $5\Gyr$ ago the Milky Way also had enough to proceed for only $2 \Gyr$ but somehow it went on for much longer, because it has accreted new gas.}

\discuss{Chao Liu}{Does the galactic fountain pollute the corona with heavy elements?}

\discuss{Answer}{Yes, this is one of the reasons why the cooling time of the corona decreases dramatically in the region of influence of the fountain causing its prompt cooling and accretion.}

\discuss{Hans-Walter Rix}{Can you say specifically how GAIA and LAMOST will test your corona-cooling scenario?}

\discuss{Answer}{GAIA and LAMOST will provide crucial constraints for chemical evolution models of the Galactic disc. These models often treat gas accretion as a free parameter, instead our supernova-driven cooling can provide an independent and self-consistent way to include accretion in the picture. For instance, our model can predict the density of pristine gas infall as a function of Galactic radius given a certain SFR density; essentially in this scheme the accretion is constrained once other properties of the Galaxy are known such as its SFR density and virial mass.}

\discuss{Yun-feng Chen}{If there are disc-corona interactions, then can we expect that the mixture is more efficient in spiral arm regions?}

\discuss{Answer}{Yes, our model is, at the moment, axisymmetric but an obvious improvement would be to include the effect of spiral arms \citep[see e.g.,][]{Struck&Smith09}.}


\end{discussion}

\end{document}